# Boson sampling with 20 input photons in 60-mode interferometers at $10^{14}$ state spaces

Hui Wang[1,2], Jian Qin[1,2], Xing Ding[1,2], Ming-Cheng Chen[1,2], Si Chen[1,2], Xiang You[1,2], Yu-Ming He[1,2], Xiao Jiang[1,2], Z. Wang[3], L. You[3], J. J. Renema[4], Sven Höfling[5,6,2], Chao-Yang Lu[1,2], Jian-Wei Pan[1,2]

[1] Hefei National Laboratory for Physical Sciences at Microscale and Department of Modern Physics, University of Science and Technology of China, Hefei, Anhui 230026, China.
[2] CAS Centre for Excellence and Synergetic Innovation Centre in Quantum Information and Quantum Physics, University of Science and Technology of China, Hefei, Anhui 230026, China.
[3] State Key Laboratory of Functional Materials for Informatics, Shanghai Institute of Microsystem and Information Technology (SIMIT), Chinese Academy of Sciences, 865 Changning Road, Shanghai 200050, China.
[4] Adaptive Quantum Optics Group, Mesa+ Institute for Nanotechnology, University of Twente, PO Box 217, 7500 AE Enschede, The Netherlands.
[5] Technische Physik, Physikalisches Instität and Wilhelm Conrad Röntgen-Center for Complex Material Systems, Universitat Würzburg, Am Hubland, D-97074 Würzburg, Germany.
[6] SUPA, School of Physics and Astronomy, University of St. Andrews, St. Andrews KY16 9SS, United Kingdom.

## Abstract

**Quantum computing experiments are moving into a new realm of increasing size and complexity, with the short-term goal of demonstrating an advantage over classical computers. Boson sampling is a promising platform for such a goal, however, the number of involved single photons was up to five so far, limiting these small-scale implementations to a proof-of-principle stage. Here, we develop solid-state sources of highly efficient, pure and indistinguishable single photons, and 3D integration of ultra-low-loss optical circuits. We perform an experiment with 20 single photons fed into a 60-mode interferometer, and, in its output, sample over Hilbert spaces with a size of $10^{14}$—over ten orders of magnitude larger than all previous experiments. The results are validated against distinguishable samplers and uniform samplers with a confidence level of 99.9%.**

There has been significant progress in demonstrating the fundamental building blocks of quantum computers (*1-11*) and quantum algorithms (*12-15*). Beyond small-scale demonstrations, the field of quantum computing is heading toward into a new regime with increasing size and complexity, where the results cannot be efficiently simulated by classical means (*16, 17*). Such a goal was referred as "quantum computational supremacy" (*18*) and "noisy intermediate-scale quantum" technologies (*19*). To this end, experimental efforts have been devoted to increase both the quality and quantity of qubits in various physical systems (*6-11*).

Boson sampling (*20*) is considered as a strong candidate for demonstrating the quantum computational supremacy. It is performed by sending $n$ identical bosons into an $m$-mode ($m>n$) Haar-random interferometer, and sampling the output distribution in the photon number basis from the output Hilbert space of the final state. Due to the bosonic statistics, the probability amplitudes of the final state are related to the permanent of submatrices of the matrix $U$ which describes the interferometer. It is strongly believed that a moderate-size boson-sampling machine will be intractable to be simulated with state-of-the-art classical computers (*21*). It is believed that the first application of quantum supremacy will be the generation of verified random numbers (*22,23*).

So far, all implementations of boson-sampling, using parametric down-conversion (*6, 24-32*) or quantum dots (*33-35*), involved at most up to five single photons and 16 interferometric modes. In those proof-of-principle experiments, the full output photon distribution was easily calculated and could be completely verified, with even the earliest classical computers. An important goal is to scale up the boson sampling into a new, computationally interesting, regime. To this end, the roadmap is to construct multiphoton boson-sampling machines with increasingly larger photon and mode number, and faster sampling rates. For a boson sampler that is large enough to demonstrate a quantum advantage, the possible number of outputs will be so large that the output samples will be sparse, i.e. each output will only be observed once in any reasonable experiment. The size of the output Hilbert spaces, which is one of the manifestations of the highly complexity, high-entropy nature of boson sampling, is a function of both the photon and mode number and can be calculated by binomial $(m+n-1, n)$. The size of the Hilbert space also determines the number of random bits generated by a verified random number generator. The output state spaces ranged from 20 to 15504 in the previous experiments (*6, 24-35*), which all the possible output events were collected for proof-of-principle studies. But those sizes are still much smaller from the actual sampling regime in which quantum supremacy can be demonstrated.

In this work, we scale up the boson-sampling with 20 photons injecting into a 60-mode interferometer where the output Hilbert space reaches $10^{14} \sim 2^{48}$, which is over ten orders of magnitude larger than the previous work. In such an exponentially large Hilbert space, for the first time, it becomes impossible in a boson-sampling experiment to exhaust all possible output combinations. Using the quantum machines constructed in this work, the achieved sampling rate becomes classically comparable to the computational power of a medium-scale integrated circuit, while a full verification by calculating the whole probability distribution will take hours using supercomputers.

We start by describing the experimental set-up of our boson-sampling machines which is illustrated in Fig. 1 for an overview. Pulsed single photon streams are produced from an InAs/GaAs quantum dot cooled to 4 K and deterministically coupled to a micropillar cavity (*36, 37*). Under pulsed resonant laser excitation (*38*), a clear Rabi oscillation of the resonance fluorescence single-photon counts as a function of pump power is shown in Fig. 2A. At π pulse with a repetition rate of 76 MHz, we eventually detect ~16 million counts per second single-mode fibre-coupled photon counts on a superconducting nanowire single-photon detector with an efficiency of 82%. The single-photon purity of the solid-state source is characterized by Hanbury Brown and Twiss measurements which reveal a second-order correlation of 0.025(1) at zero-time delay (Fig. 2B), indicating a single-photon purity of 97.5%. The photon indistinguishability is then measured by Hong-Ou-Mandel interferometers with two photons' emission time separation up to ~6.5 μs. The measured photon indistinguishability is 0.954(1) at a short time separation of ~13 ns, which slightly drops to a plateau of 0.923(1) between ~1 μs and ~6.5 μs (Fig. 2C). Such a semiconductor source of polarized, high-brightness, high-purity, and near-transform-limited single photons is the central quantum resource for boson sampling.

The single-photon stream is then deterministically de-multiplexed into 20 spatial modes using fast optical switches, arranged in a tree-like structure (see Fig. 1). Each switch consists of a polarizing beam splitter (with an extinction ratio >2000:1) and an active Pockels cell that is synchronized to the laser pulses and on-demand rotates the photon polarization (with an extinction ratio >100:1). The measured average system efficiency of the switches is ~83% (*39*). After finely compensating for their relative time delay, these 20 demultiplexed modes are fed into a fully connected 60-port linear optical network. Finally, the 60 output ports are fed into 60 superconducting nanowire single-photon detectors with their efficiencies varying from 60% to 82%.

Multi-mode interferometers are usually constructed by beam splitters and phase shifters in a triangular (*6, 40*) or rectangular (*34, 41*) configuration. Here, we put forward a new, more compact 3D design that combines phase stability, full connectivity, matrix randomness, near-perfect wave-packet overlap, and near-unity transmission rate simultaneously (see Fig. 3A and 3B and its caption). Such an optical network consists of 396 beam splitters and 108 mirrors, and can be used to implement 60×60 unitary transformations. We use Mach-Zehnder type interferometry to calibrate spatial overlap between any two input ports, which shows an average visibility above 99.9%. From all the 20 input ports, the transmission rate of the whole optical network is measured to be 98.7%, and the average coupling efficiency in all the 60 output ports is ~90%.

We use a narrowband laser to reconstruct the corresponding unitary matrix of the 3D 60×60 interferometer. The measured elements of the amplitudes and phases are shown in Fig. 3C and 3D, respectively. If the generated matrix is unitary, the product of this matrix and its Hermitian conjugate should be an identity matrix. The result is plotted in Fig. 3E showing that the average of the non-diagonal elements is 0.01, thus confirming the high degree of unitarity of the generated matrix. Moreover, for the hardness arguments regarding boson sampling to hold, the matrix should be randomly drawn according to the Haar measure. We compare our measured elements with ideal Haar-random matrix elements. Figure 3E (3F) shows the statistical frequency of the measured 1200 elements of amplitude (phase), which reasonably agrees with the predication from Haar-random matrix.

We use fidelity ($F$) and total variation distance ($D$) to quantify the performance of the boson-samplers, which are defined by: $F = \sum_i \sqrt{p_i q_i}$, and $D = \sum_i |p_i - q_i|/2$ ($p_i$ and $q_i$ denote the theoretical and experimental probability of $i$-th basis, respectively). For a perfect boson-sampler, the fidelity should equal to 1 and the distance should be 0. To test the boson sampling set-up works properly, we first analyze arbitrary two-photon input configurations, of which there are 190 in the 20-input set-up. We obtain ~300,000 samples for each configuration (~170 times larger than the state spaces). The measured fidelities and distances between the experimental and ideal cases are illustrated in Fig. 4A, from which an average fidelity of 0.995(3) and distance of 0.043(5) are extracted. If we use the Clifford-Clifford sampling algorithm (*42*) to generate the same number of samples on a classical computer, the total variation fidelity (distance) is 0.998 (0.035), which is only slightly better than those from the quantum machine. This indicates high level of interference between any two modes in the 60-mode interferometer. The small error in the two-photon test is mainly from the finite sample number and statistics. Next, we register the whole output distributions of the non-collision events from three- and four-photon boson sampling, which are plotted in Fig. 4B and 4C. The measured fidelities and distances are 0.988(1) and 0.095(1) for the former, and 0.984(1) and 0.143(1) for the latter. These results are in excellent agreement with the theory taking into account of the realistic single-photon source and optical network (*39,43-47*), which confirms that the boson-sampler works properly.

As the photon number increases, the output Hilbert spaces expand exponentially, which makes it infeasible to register the whole distribution—but only "sampling" is possible, a regime long waited for in boson-sampling experiments. Meanwhile, due to the passive nature of the boson-sampling protocol, the output multi-photon coincidence rate drops exponentially. In our experiment, when the photon number exceeds four, the registered distributions become sparse, that is, most of the output combinations record zero events.

We operate both standard Aaronson-Arkhipov model (*20*) and Aaronson-Brod model of boson-sampling (*48*) with photon loss. In the former, all the *n* input photons are detected in the output. In the latter, *n+k* photons are sent in, but in the output only *n*-fold coincidences are detected. The sampling rate of the latter is enhanced by a factor of approximately $\text{binomial}(n+k, n)$ compared to the former. In this work, we perform the standard model of boson-sampling for coincidence detection of 10 photons or below, and the lossy boson-sampling in the regime of more involved photons—up to 20 input and 14 detected photons. As plotted in Fig. 4D, the coincidence rate in the standard boson-sampling is 295 Hz for five photons, which is 60 times higher than in (*6*), and drops to 0.01 Hz for 10-photon coincidence. Tolerating one photon loss can give rise to an approximately 30 times enhancement in the sampling rate, as plotted in Fig. 4D. For 20 input photons with six lost, we detect 14-photon coincidence rate of ~6 per hour, which allows us to obtain 150 samples after a collection time of 26 hours. In all settings, we obtain at least a few hundreds of samples to characterize our multiphoton boson-sampler.

An exponentially large output state space in the boson sampling represents an important manifestation of its complexity. Using the boson sampling parameters in this experiment, we calculate the state spaces and plot them in Fig. 4E. While the previous experiments were limited to state spaces between 20 and 15504, our work achieves Hilbert space dimensions up to $\sim 3.7 \times 10^{14}$ in the 20-photon-input 14-output boson-sampling, which is more than ten orders of magnitude larger than before. With such enormous output state spaces, it is no longer possible to collect the full distribution as in the previous small-scale experiments (*6,24-35*). In fact, theoretically calculating the full probability distribution in the 20-photon-input 14-output boson-sampling will take hours using supercomputers.

While full verification of large-scale boson-sampling is also strongly conjectured to be intractable for classical computation, there are methods for validating boson-sampling that can provide supporting or circumstantial evidence for the correct operation of this protocol. We use a statistical test to first rule out a possible hypothesis that the input photons are distinguishable, which is very relevant to the experimental implementations because single photon's indistinguishability is most susceptible to the decoherence. We perform Bayesian analysis (*49*) for various input-output configurations. Typical results with 14-20 input photons are plotted in Fig. 5A, showing an increasing difference between indistinguishable bosons (solid) and distinguishable bosons (hollow). With ~50 samples, these analyses already reach confidence levels of ~99.9% that these results are from indistinguishable bosons. The validation data for other photon numbers can be found in (*39*). Second, we aim to rule out uniform distribution, that is, the samples scatter uniformly to the overall distribution. We employed the row-norm estimator test (*50*), where the increasing difference between experimental data (solid) and simulated uniform samples (hollow) indicates that our results cannot be reproduced by a uniform sampler (Fig. 5B). We hope our experiment will inspire new theoretical methods for quantitative characterizations for large-scale boson-sampling using only very sparse samples, for example, to estimate the fidelity and total variation distance in the future.

Our results show that we can experimentally access quantum states of 20 photons in a 60×60 interferometer, and use it to perform a quantum computational task increasingly difficult for classical computers with growing number of photons. One way to bring the current work in relation with other general photonic qubit experiments (*7,51,52*) is the following. The size of the involved Hilbert space scales with $d^n$, where d stands for the dimensions and $n$ is the number of involved particles. In this regard, the Hilbert size of our experiment platform is $60^{20} \sim 2^{118}$, an unprecedentedly large size that is many orders of magnitude larger than all previous work (*6-11*). The mode dimension of the 3D interferometer in our design can be directly doubled by using spatial-polarization encoding, and can be easily scaled up to a few hundreds. With ongoing improvements of single-photon source efficiency (*53*), our experimental approach points a way to the noisy intermediate-scale quantum regime through boson-sampling.

**Figure caption:**

**Figure 1: Experimental set-up of boson-sampling.** A single InAs/GaAs quantum dot, resonantly coupled to a microcavity yielding a Purcell factor of ~18, is used to create pulsed resonance fluorescence single photons. For demultiplexing, 19 pairs of Pockels cells (PCs) and polarized beam splitters (PBSs) are used to actively translate a stream of photon pulses into 20 spatial modes. Optical fibers with different lengths are used to compensate time delays. The 20 input single photons are injected into a 3D integrated, 60-mode ultra-low-loss photonic circuit (see Fig. 3 for more details), which consists of 396 beam splitters and 108 mirrors. Finally, the output single photons are detected by 60 superconducting nanowire single-photon detectors with efficiencies ranging from 0.6 to 0.82. All coincidence are recorded by a 64-channel coincidence count unit (not shown).

**Figure 2: Characterization of the single-photon source. A.** By varying the amplitude of pumping laser field, a Rabi oscillation up to $4\pi$ is observed. At $\pi$ pulse, ~16.3 million single photons per second are recorded by a superconducting nanowire single-photon detector. **B.** Characterization of single-photon purity using second-order correlation function. The strongly antibunched peak at zero-time delay reveals $g^2(0) = 0.018(1)$. **C.** Measurement of photon indistinguishability by Hong-Ou-Mandel interference between two photons with different emission time separations. The extracted photon indistinguishability are 0.954(1), 0.948(1), 0.933(1), 0.929(1), 0.922(1) and 0.923(1) at emission time separations of 13 ns, 39 ns, 210 ns, 395 ns, 1.8 μs and 6.5 μs, respectively. The data are fitted by a model considering Markov noise.

**Figure 3: Construction and characterization of the 60-mode photonic network. A.** The 60-mode interferometer consists of one rectangular piece and two triangular pieces. The rectangular is fabricated by bonding (through inter-molecular force) six trapezoidal pieces with a size of 28.28×28.28×4.00 mm$^3$. The triangular is constructed in a similar way, with a size of 24.00×24.00×20.00 mm$^3$. The bonding interfaces are coated with random polarization-dependent beam-splitting ratio. The trapezoidal pieces are cut and bounded with a dimension tolerance of <5 μm and parallel precision better than 5''. The design ensure that any possible spatial mismatch is much smaller than coherence length of the quantum-dot single photons (~30 mm). **B.** Illustration of light propagation inside the 3D photonic circuit. The rectangular piece has six horizontal 10-mode layers, while the two triangular ones have 10 vertical six-mode layers. In the rectangular piece, only the photons in the same horizontal layer can interfere with each other, but not with vertically different layers. After that, ten vertical layers are incorporated, which are to make the photons from different horizontal layers to interfere with each other. Therefore, the interferometer fully connected. **C.** The measured 1200 elements of amplitude. These values are determined by the recorded counts of the 60 single-photon detectors when we inject photons in every input port one by one. **D.** The measured 1200 elements of amplitude. These elements are measured using Mach-Zehnder type interference with a narrowband laser. **E.** Unitarity test of the reconstructed matrix. All output probability are normalized to unity, corresponding to 20 diagonal elements. The average of all off-diagonal elements is as small as 0.01, confirming that the matrix is well reconstructed. **F.** Statistical histogram of 1200 elements of amplitude. **G.** Statistical histogram of 1200 elements of phase. The phase is uniformly distributed from -π to π.

**Figure 4: Experimental results of boson-sampling. A.** Summary of distance (purple) and fidelity (orange) of all 190 possible two-photon boson-sampling tests. The averaged distance and distance are 0.043(7) and 0.996(1), respectively. **B.** Reconstructed three-photon distribution for an input combination (2,10,12). **C.** Reconstructed four-photon distribution for an input combination (1,2,4,7). **D.** Boson-sampling rates with different settings of input photons and lost photons. **E.** Summary of the dimensions of output Hilbert space of all previous work and this experiment (red dots).

**Figure 5: Validation of the boson-sampling results. A.** Bayesian analysis of typical 11- to 14-photon boson sampling results. The solid points are from Bayesian analysis by updating every experimental data, which reach a level of 99.9% with only ~50 samples. These results confirm that the experimental events are from genuine boson sampler, rather than distinguishable sampler. **B.** Designed row-norm estimator to discriminate experimental data from a uniform sampler. The solid dots are updated by the experimental data, while the open dots are from simulated uniform sampler. The increasing differences between them can exclude the uniform sampling hypothesis. For clarity, in each panel, the four different data sets are displayed with an offset in x axis.

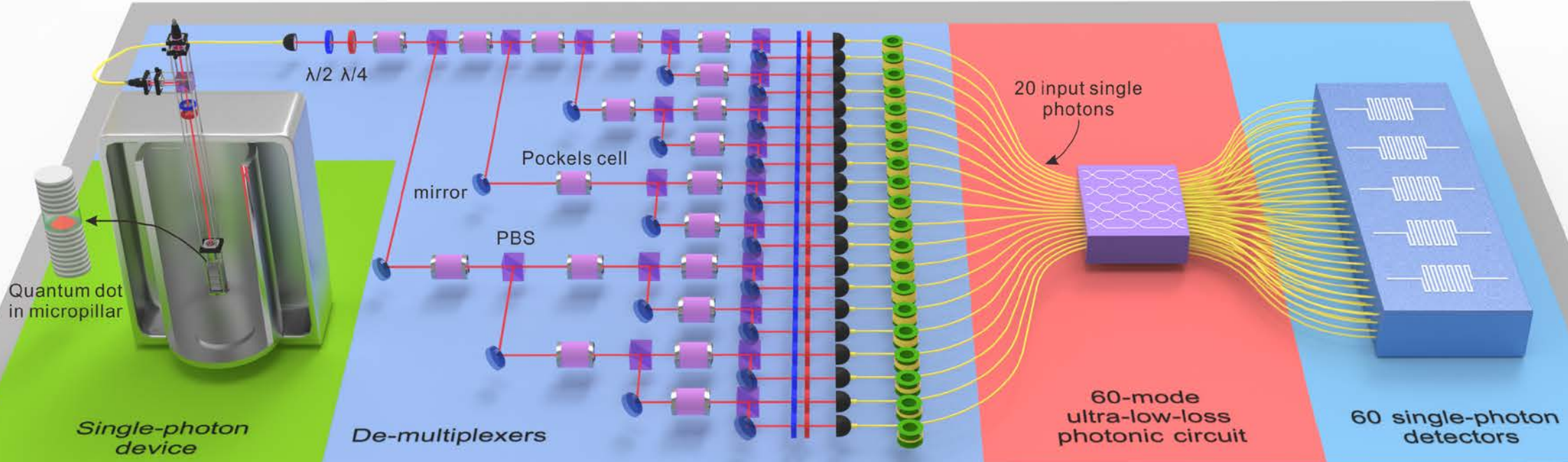

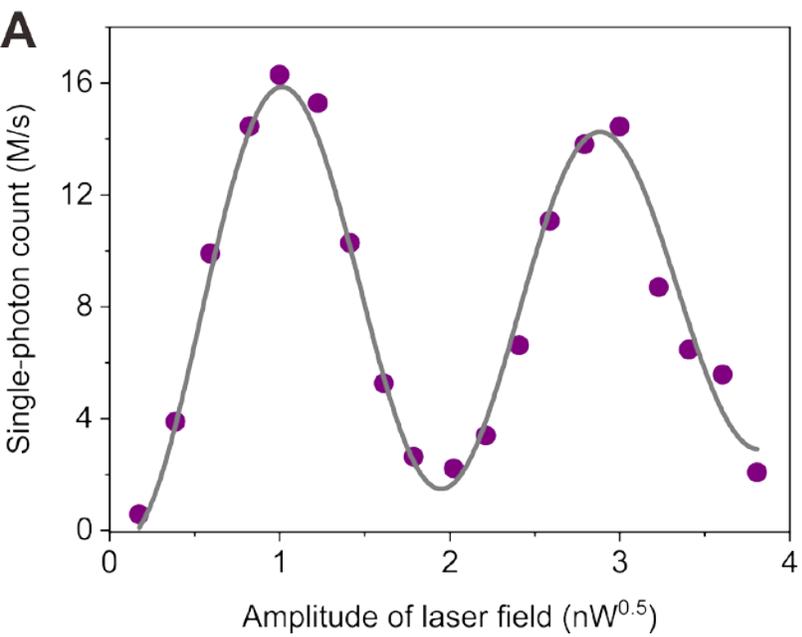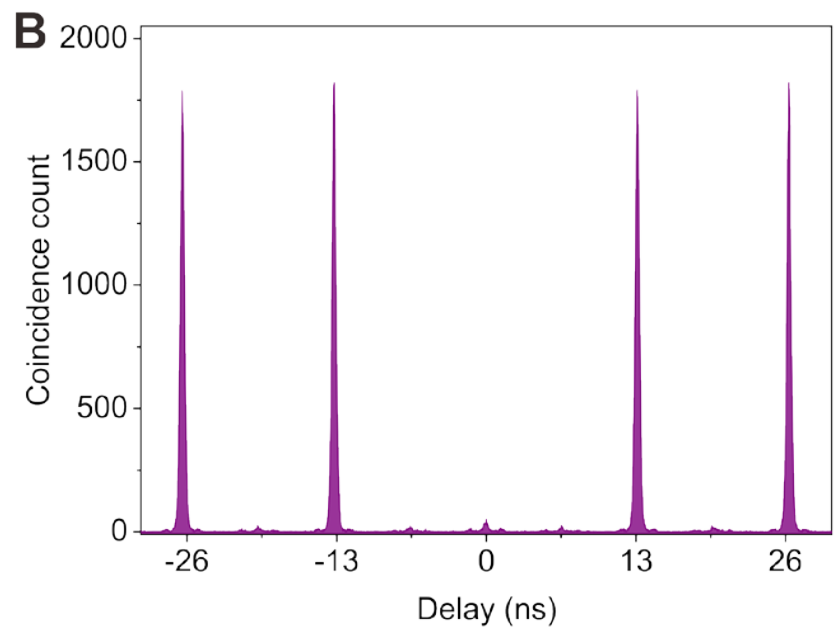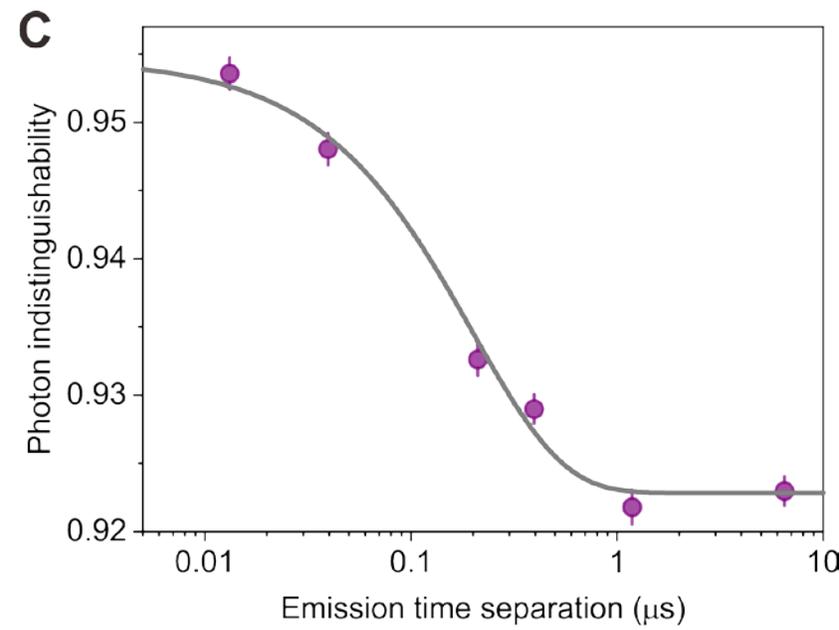

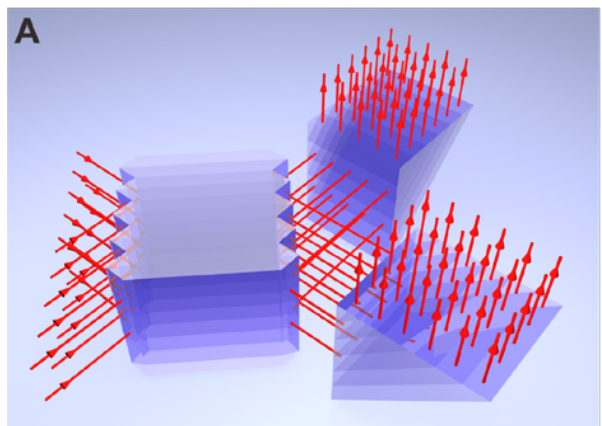
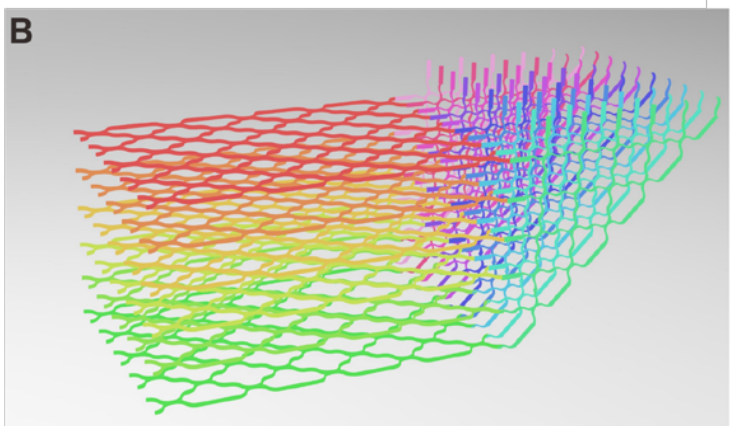
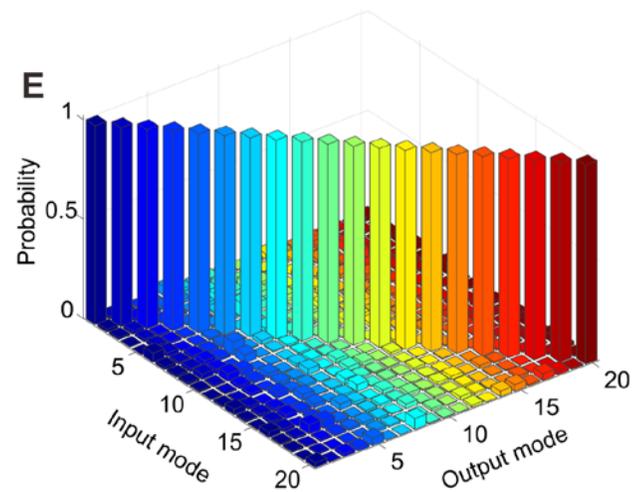
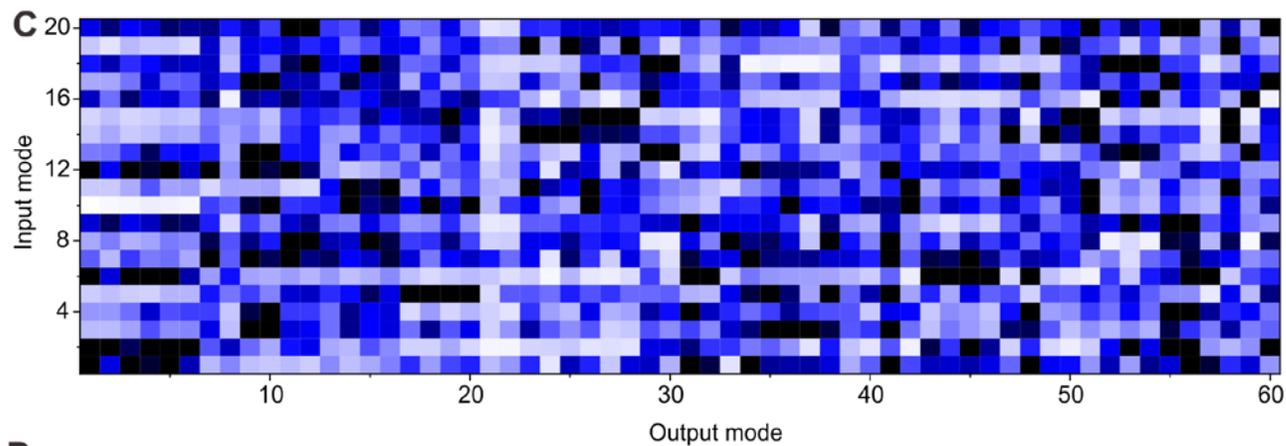
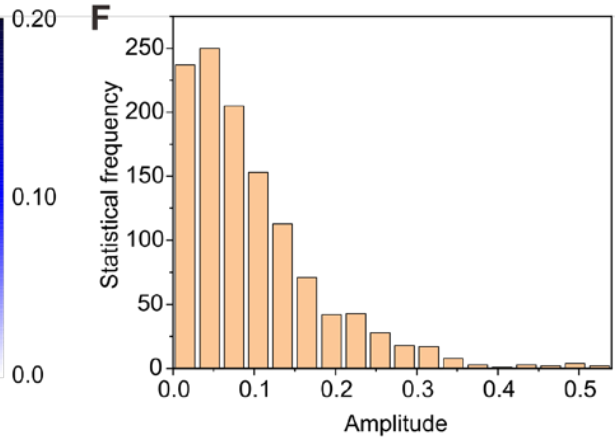
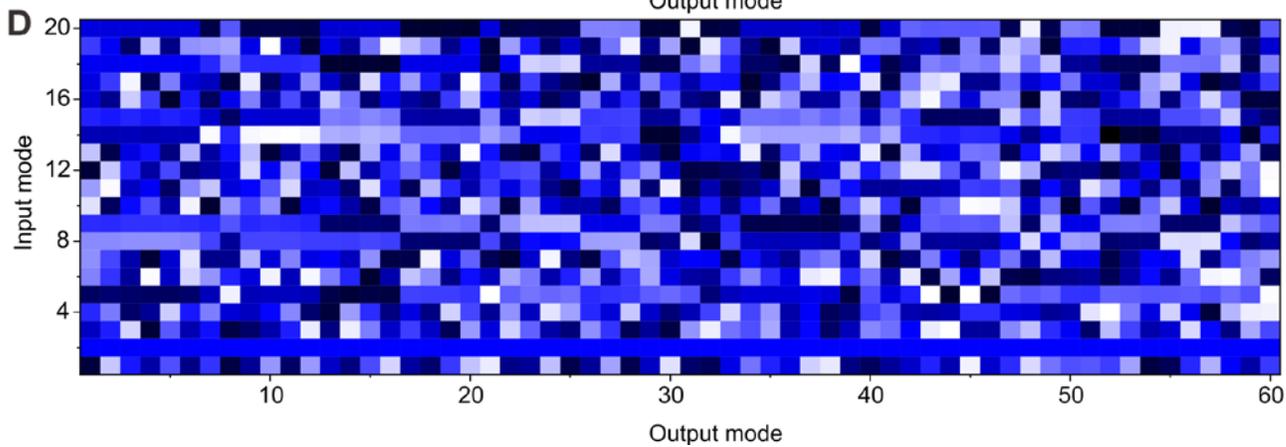
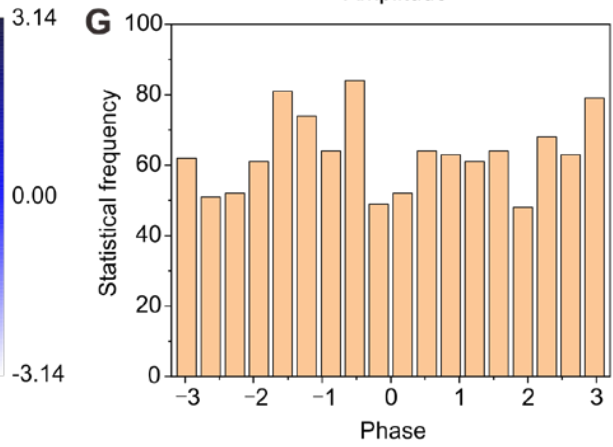

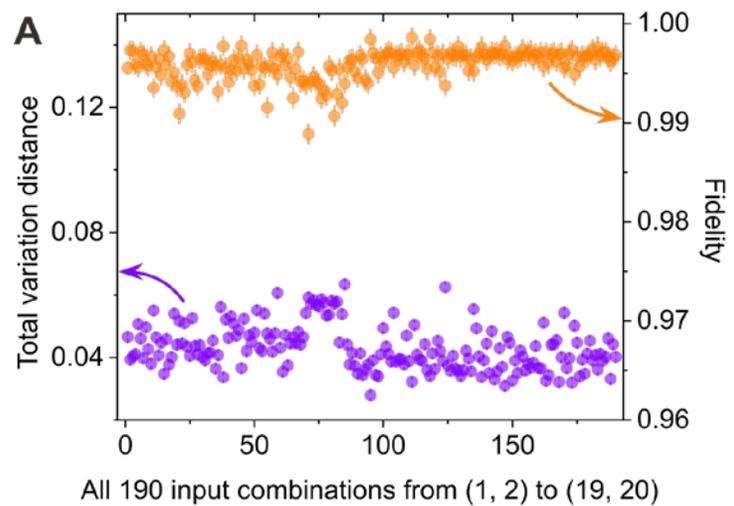
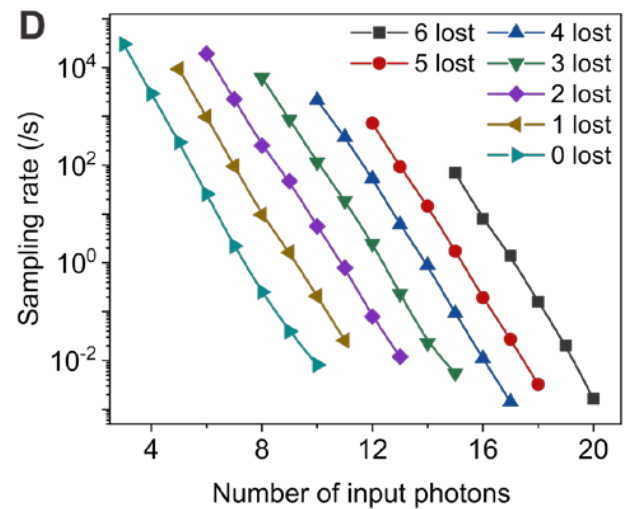
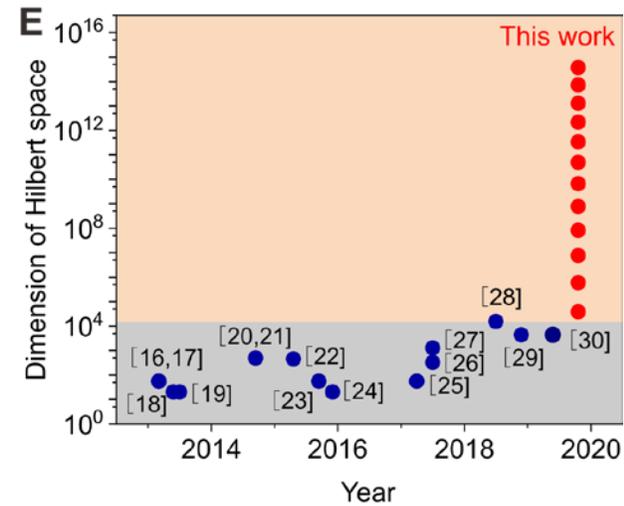
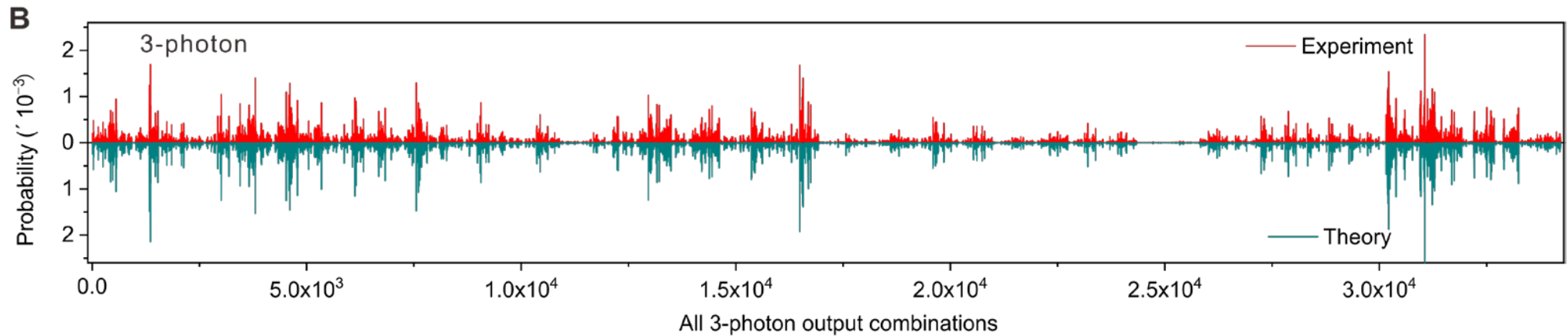
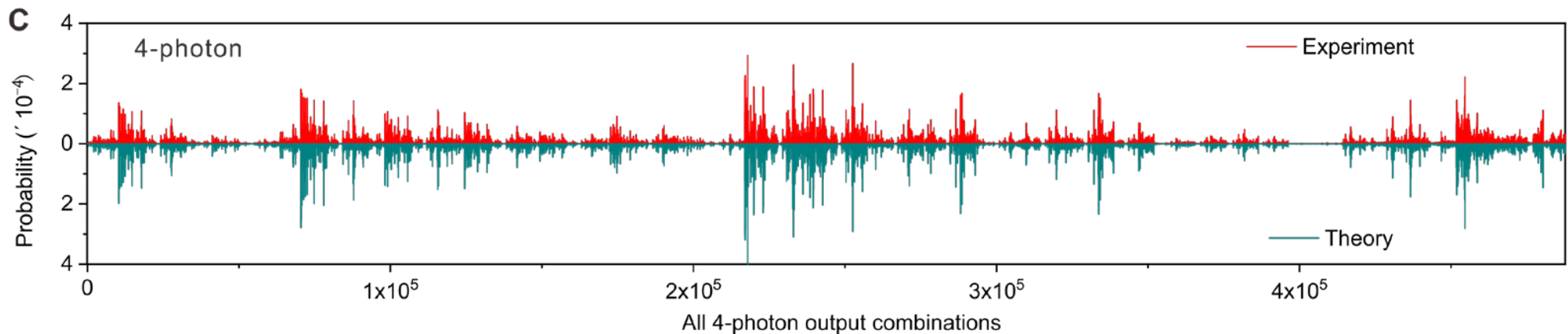

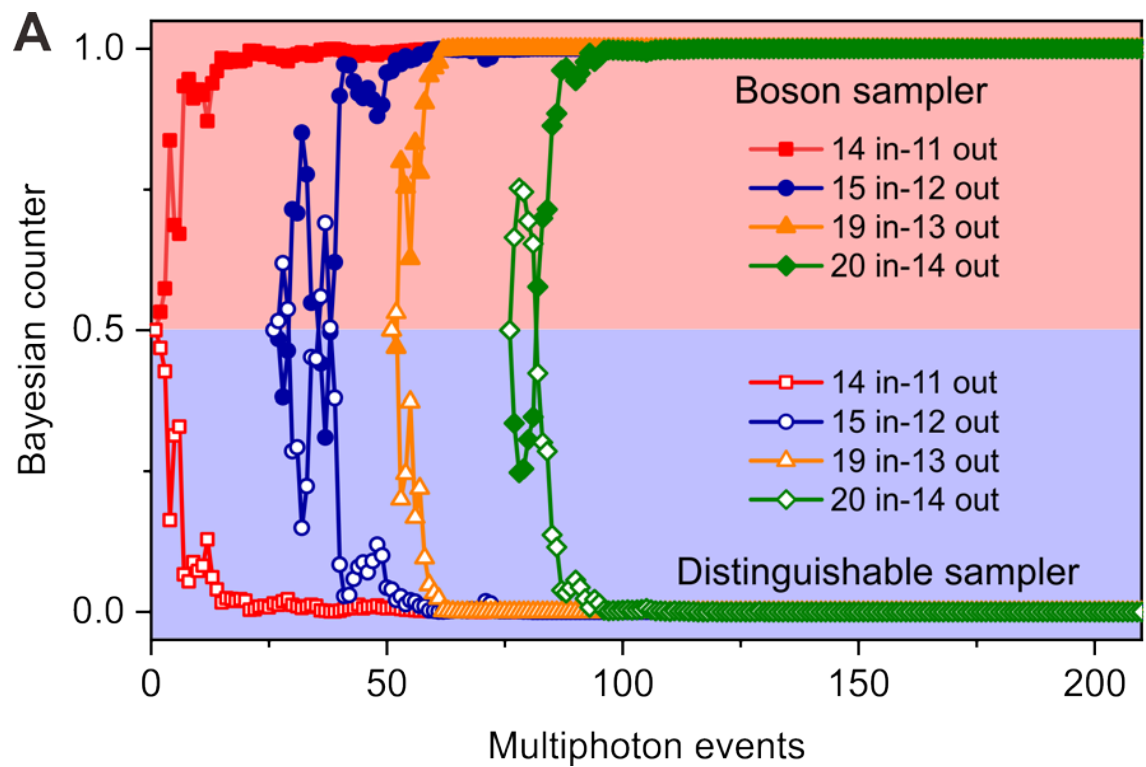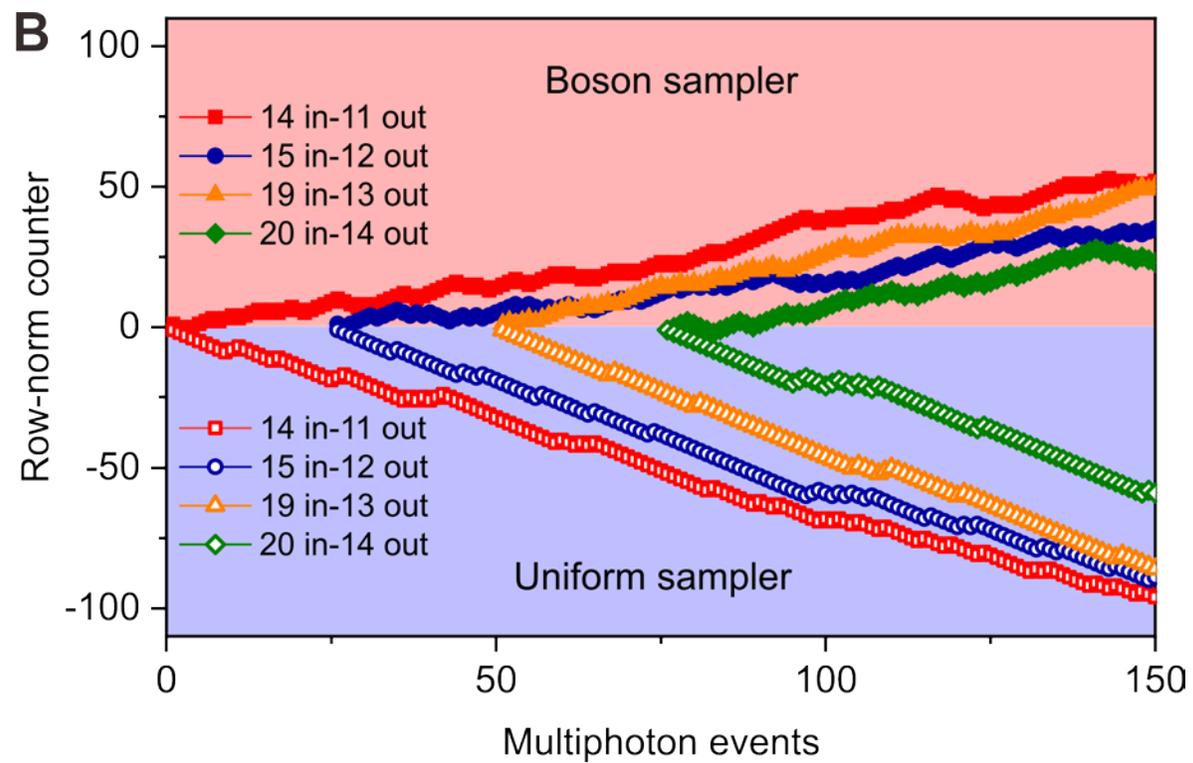

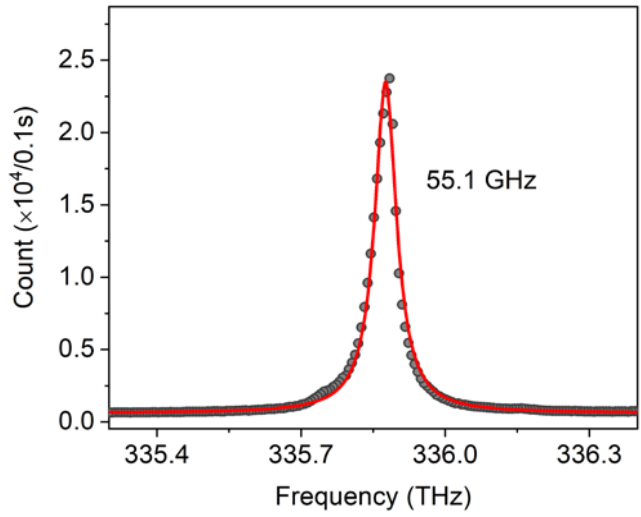

**Fig. S1:** The measured cavity mode of micropillar. The quality factor is ~6800.

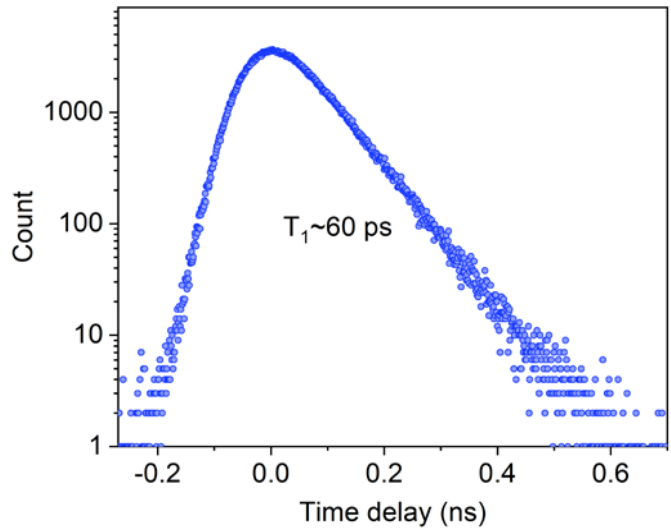

**Fig. S2:** The resonance fluorescence lifetime when the quantum dot is resonant with the micropillar cavity at ~4 K.

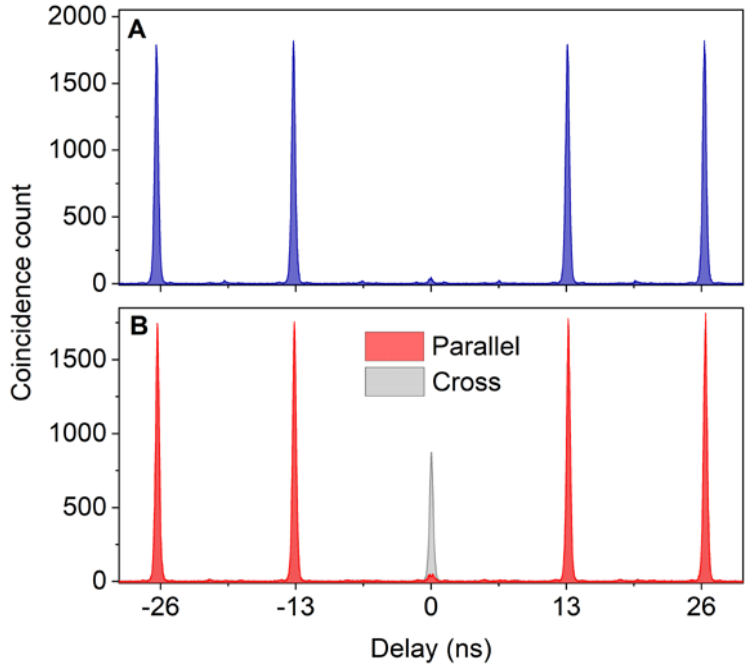

**Fig. S3:** The purity and indistinguishability of single-photon source. **A**. The second-order correlation function at zero-time delay is 0.025(1). **B**. Hong-Ou-Mandel visibility of 0.933(1) is measured with a 16-pulse emission time separation.

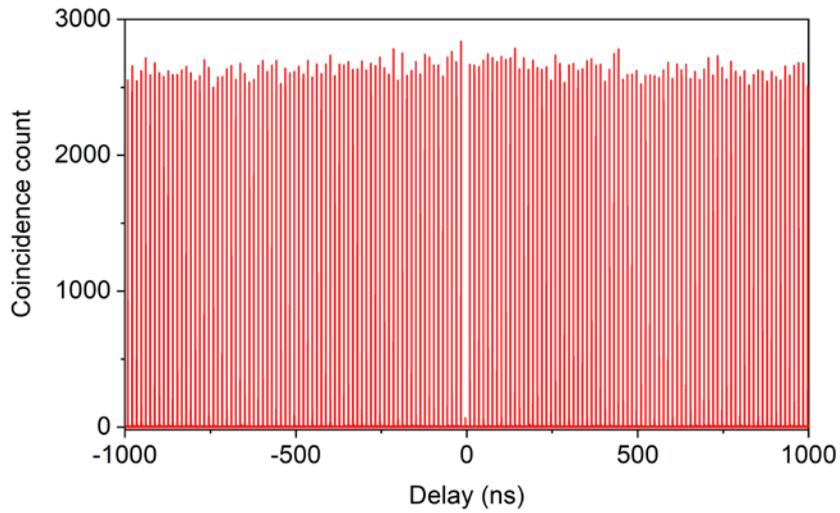

**Fig. S4:** The second-order correlation measurement at a time scale from -1000 ns to 1000 ns. The data shows that almost no blinking happens in quantum dot sample.

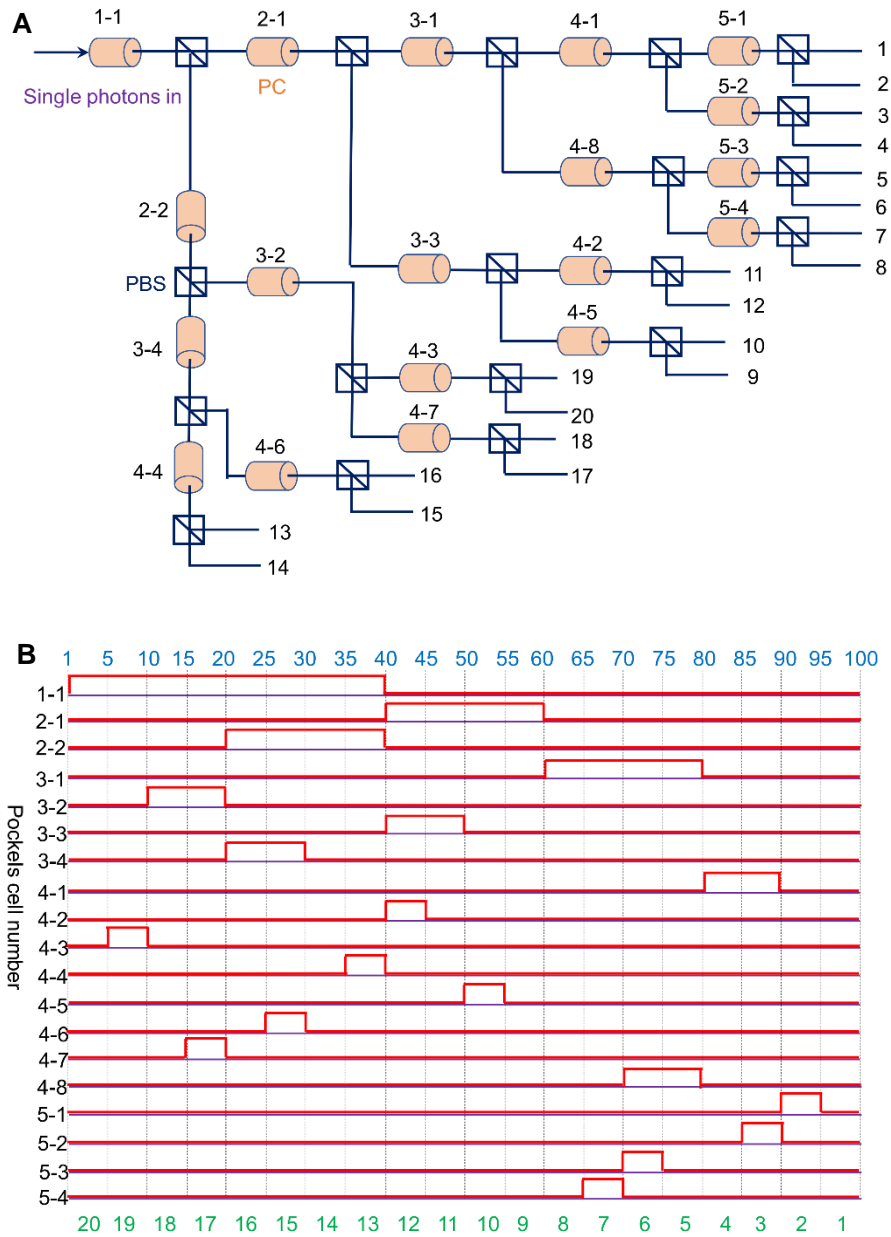

**Fig. S5: A.** The architecture of the demultiplexers used in this experiment. A stream of single photon pulses can be divided into 20 segments by 19 pairs of PC and PBS. **B.** All electric signals loaded to corresponding PCs. The width and delay for all signals must be well controlled to ensure correct operation.

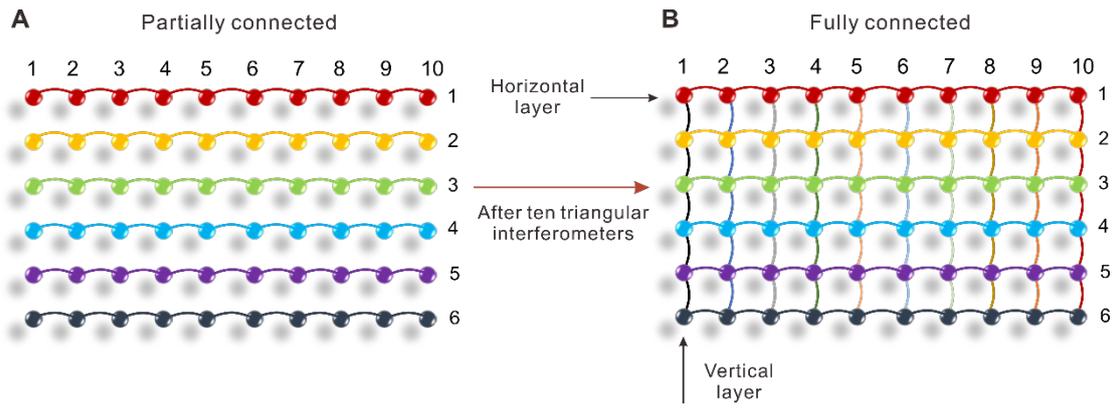

**Fig. S6:** An illustration of the interferometer. **A.** After the rectangular interferometer, all horizontal layers are connected, however, no interaction among vertical layers. **B.** After triangular interferometers, all 60 modes are fully connected.

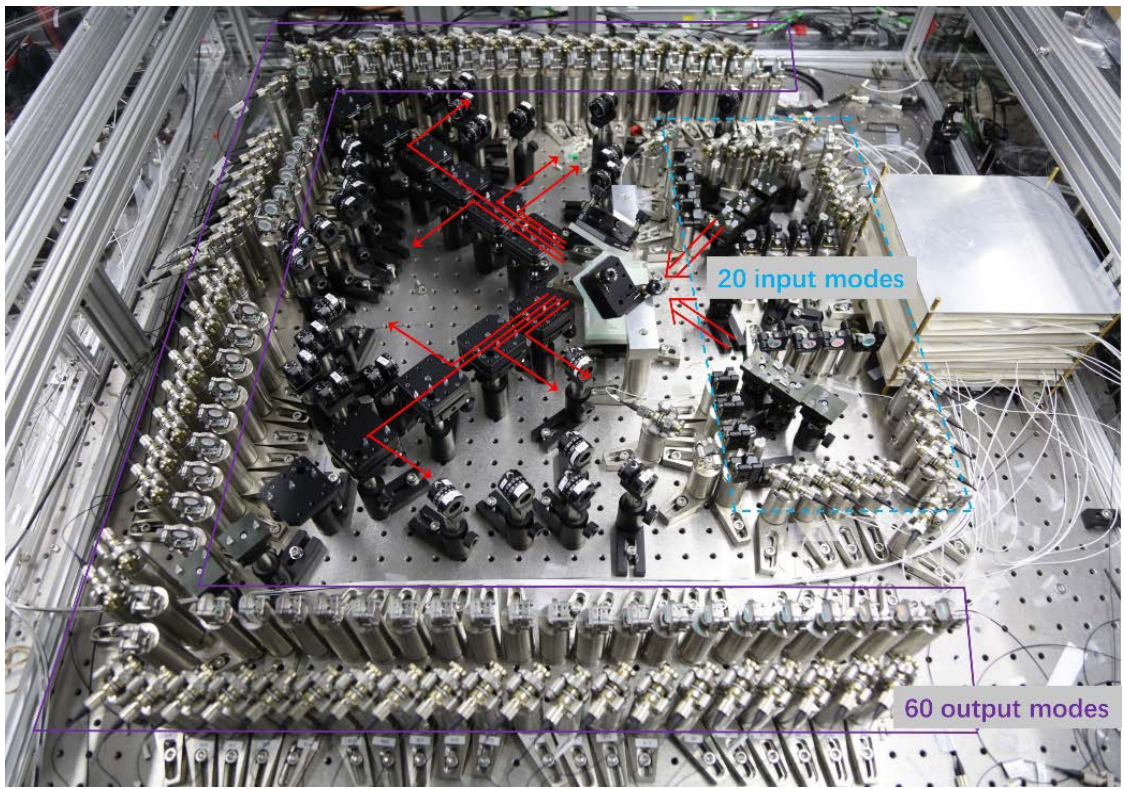

**Fig. S8:** An illustration of photonic network setup build in our laboratory. The interferometers are placed at the center of the optical table. At the right side, 20 input collimators (shown in the blue box) inject 20 beams into the photonic network. When passing through it, 60 output beams are separated by mini-mirrors to surrounding 60 collimators (shown in the purple box) to collect photons using single-mode fibers.

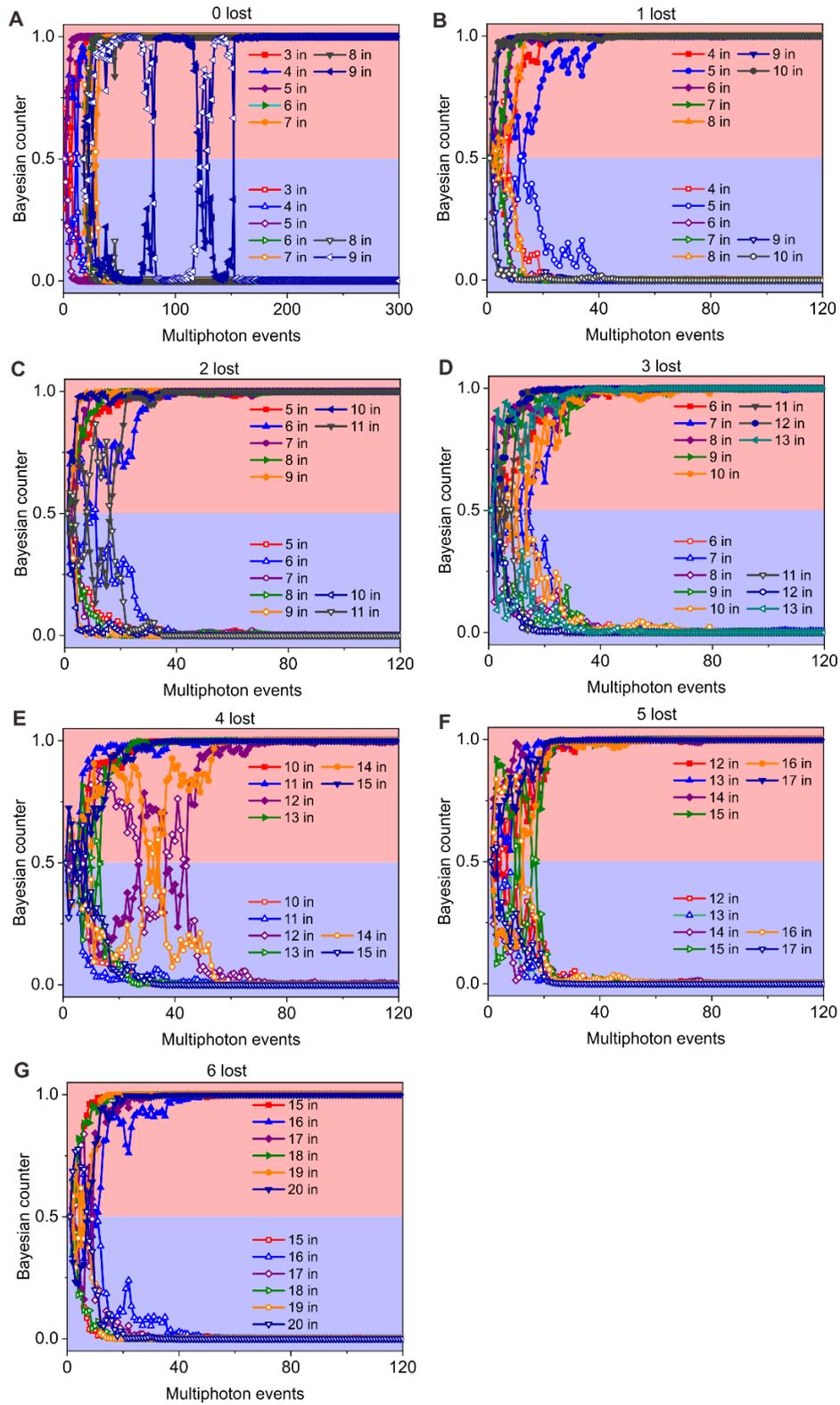

Fig. S10: Validating experimental data with Bayesian analysis to rule out distinguishable sampler. All solid dots from A to G is updated by our experimental data, where all points converged to 1 quickly, which strongly indicate that our experimental results are from boson sampling.

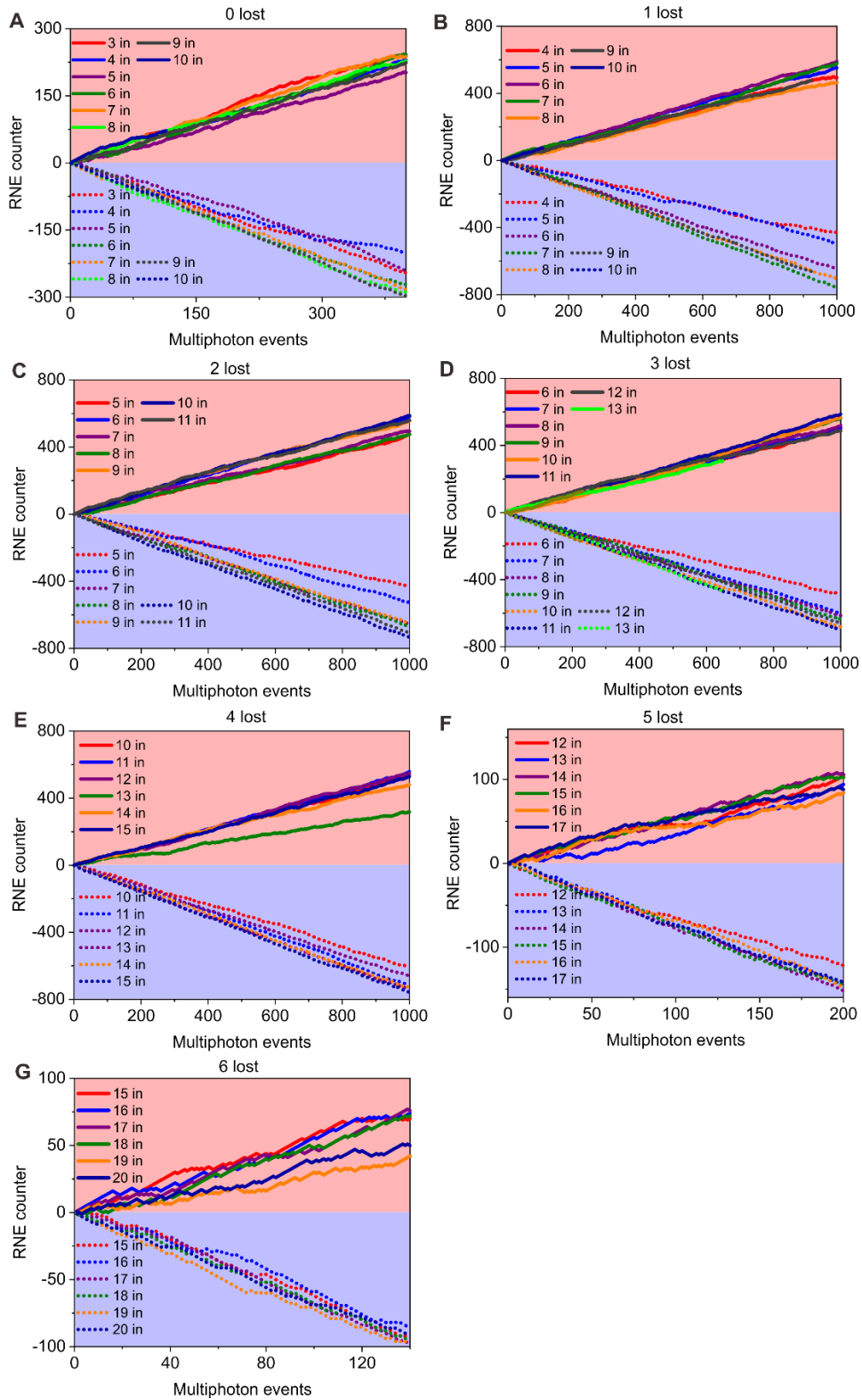

Fig. S11: Validating experimental data with RNE to rule out uniform sampler. All solid dots from A to G are updated by our experimental data, while the open dots are from simulated uniform samplers. the increasing difference between them demonstrate that boson sampling is far from uniform.